\documentclass[aps,rmp,epsf,twocolumn]{revtex4}
\usepackage{graphicx}% 
\newcommand{\ac}{\mbox{\boldmath$a$}}

\newcommand{{\vv}}{\mbox{\boldmath$v$}}
\newcommand{{\xx}}{\mbox{\boldmath$x$}}

\newcommand{\FF}{\mbox{\boldmath$F$}}
\newcommand{\XX}{\mbox{\boldmath$X$}}
\newcommand{\DD}{\mbox{\boldmath$D$}}
\newcommand{\uu}{\mbox{\boldmath$u$}}

\newcommand{\BB}{\mbox{\boldmath$B$}}
\newcommand{\CC}{\mbox{\boldmath$C$}}

\newcommand{\bDelta}{\mbox{\boldmath$\Delta$}}
\newcommand{\UNIT}{\mbox{\boldmath$1$}}
\newcommand{\bnabla}{\mbox{\boldmath$\nabla$}}

\begin{document}

\title{Role of the H Theorem in Lattice Boltzmann Hydrodynamic Simulations}

\author{Sauro Succi}
\altaffiliation[Corresponding author]{}
\email{succi@iac.rm.cnr.it}
\affiliation{Istituto Applicazioni Calcolo, CNR, viale Policnico 137, 00161 Rome, Italy}

\author{Iliya V.\ Karlin}
\email{ikarlin@ifp.mat.ethz.ch}
\affiliation{ETH-Z\"{u}rich, Institute for Polymer Physics, CH-8092 Z\"{u}rich, Switzerland}

\author{Hudong Chen}
\affiliation{EXA Corporation, 450 Bedford street, Lexington, Massachusetts 02420, USA}

\begin{abstract}

In the last decade, minimal kinetic models, and
primarily the Lattice Boltmann equation, 
have met with significant success in the simulation of
complex hydrodynamic phenomena, ranging from slow flows
in grossly irregular geometries to fully 
developed turbulence, to flows with dynamic phase transitions.
Besides their practical value as efficient computational 
tools for the dynamics complex systems, these minimal models may 
also represent a new conceptual paradigm in modern 
computational statistical mechanics: 
instead of proceeding bottom-up from the underlying 
microdynamic systems, these minimal kinetic models are built
top-down starting from the macroscopic target equations.
This procedure can provide dramatic advantages, provided
the essential physics is not lost along the way.
For dissipative systems, one such essential requirement
is the compliance with the Second Law of thermodynamics.

In this Colloquium, we present a chronological
survey of the main ideas behind the Lattice Boltzmann method, with
special focus on the role played by the $H$ theorem in enforcing
compliance of the method with macroscopic evolutionary constraints
(Second Law) as well as in serving as a numerically stable computational
tool for fluid flows and other dissipative systems out of equilibrium.

\end{abstract}
\maketitle

\tableofcontents

\section{Introduction}

The quest for a better understanding of the macroscopic world in
terms of underlying ``fundamental'' microscopic laws is a timeless
issue in the history of science and natural philosophy.
Modern science has provided an admirably powerful theory, and mathematical
tool as well, to address this issue in a sensible and productive way, a gift
named Newtonian mechanics.
Newtonian mechanics is a theory of amazing depth and breadth, going
as it does from scales of planetary motion, all the way down to molecular
trajectories, encompassing almost {\it twenty orders of 
magnitude} in the process.

Besides an array of practical results, the application of Newtonian 
mechanics at the molecular scale generates a profound puzzle: the
origin of {\it irreversibility} and the nature of time itself.
Newton's equations are manifestly reversible, that is, invariant under
time and velocity inversion, which means that molecular motion is
basically like a movie which can be indifferently rolled forwards or
backwards in time with no loss of information.
This is in blatant contrast to our daily experience of an inexorably
one-sided nature of time evolution: the arrow of time travels only one-way.
This puzzle has remained outstanding
for more than a century, and is still open to a large
extent, and in any case is beyond the scope of the present work.
Besides its profound philosophical implications, irreversibility
also bears upon the practical question of predicting the macroscopic
behaviour of complex systems comprising a huge number of (non-linearly)
interacting individual units.
The application of Newtonian mechanics to the molecular scale
inevitably leads to a gigantic wave of {\it computational
complexity} due to the enormous number of atoms/molecules constituting
macroscopic systems. This problem is circumvented by formulating
{\it continuum} models, typically based on partial
differential equations describing the space-time evolution of a few
macroscopic fields, such as fluid density, 
pressure, temperature, and so on. 
This approach is indeed quite successful, but when
confronted with complex systems out of equilibrium, e.g., fully
turbulent flows, it shows clear limitations.
The point is that, owing to strong non-linearities
and multi-dimensionality, the aforementioned partial 
differential equations are often just too complicated 
to be solved by even the most powerful numerical techniques.
It therefore makes sense to go back to Newtonian-like dynamics,
namely large sets of {\it ordinary} differential equations, and
develop {\it minimal} fictitious particle
dynamics designed so as to relinquish as many microscopic details as
possible without corrupting the ultimate macroscopic target.

In this work, we shall be concerned precisely with this 
type of modelling strategy.
In particular, we shall turn our attention to alternative ways to
gain understanding about the {\it predictability} of macroscopic 
phenomena out of hyperstylized ``Newton-like'' microscopic {\it models}. 
We hasten to add that these alternative routes are highly influenced
by advances in computational modelling, and therefore
they naturally fit into the general
framework of computational statistical mechanics. 
Like their real-life physical counterparts, these hyperstylized
models are required to display irreversible behaviour as a basic
requisite of stability, whence the importance of designing them
in compliance with the second law of thermodynamics.
In this Colloquium we shall be concerned with the Lattice
Boltzmann equation, a minimal form of the Boltzmann
equation which retains just the least amount of kinetic information
neeeded to recover correct hydrodynamics as a macroscopic limit. 
The Lattice Boltzmann equation
has proved quite effective in describing a variety
of complex flow situations using a very simple and elegant formalism
built upon the aforementioned hyperstylized approach. 

\section{Statistical mechanics background}

The theory of the Lattice Boltzmann equation
belongs to the general framework of non-equilibrium
statistical mechanics. 
In this section we shall therefore present a brief
review of the main cornerstones of classical statistical
mechanics, starting from the most fundamental (atomistic)
level, all the way up to the macroscopic level.

For most practical purposes, our ability to predict the behaviour 
of the world around us depends upon the time evolution of macroscopic variables,
e.g., pressure, temperature, flow speedsr, etc., which result from the 
collective
average over of an enormous amount of individual trajectories.
Since we can only experience average quantities, it makes sense to think
of mathematical formulations dealing directly with these average
quantities: which is the chief task of Statistical Mechanics.

\subsection{The BBGKY hierarchy}

The traditional route to this end is the celebrated
BBGKY (Bogoliubov-Born-Green-Kirkwood-Yvon) hierarchy
(Cercignani, 1975, Liboff, 1998), leading from
atomistic equations to fluid dynamic equations for
the macroscopic variables, typically the Navier-Stokes equation
of fluid flow (Landau and Lifshitz, 1953).

The BBGKY path is based on the four basic levels
(see left branch of Figure 1):
{\em
\begin{itemize}
\item Atomistic level (Newton-Hamilton)
\item Many-body kinetic level (Liouville)
\item One-body kinetic level (Boltzmann)
\item Macroscopic level (Navier-Stokes)
\end{itemize}
}

Let us discuss these four levels is some more detail.

\subsection{The atomistic level}

The atomistic description of (classical) macroscopic
systems is based on Newtonian mechanics.
The mathematical problem generated by Newtonian
mechanics is to solve a set of $N$ 
non-linear ordinary differential equations:
\begin{equation}
\label{NEWTON}
m_i\frac{d^2 {\xx}_i}{dt^2} = {\FF}_i,\\ 
\end{equation}
with the initial conditions:
\begin{eqnarray}
\label{NEWTONIC}
{\xx}_i(t=0)={\xx}_{0i}\\\nonumber
{\vv}_i(t=0)={\vv}_{0i},\;\;\;i=1,N
\end{eqnarray}
where $N$ is of the order of Avogadro's number
$N_A \sim 6 \times 10^{23}$.
In the above, $m_i$ are the molecular masses, ${\vv}_i=d{\xx}_i/dt$ the
molecular speeds and ${\FF}_i$ is the force acting upon the $i$-th 
molecule due to intermolecular interactions (Huang, 1987).

The application of Newtonian mechanics to the molecular world prompts 
a daunting wave of {\it computational complexity}.
A centimeter cube of an ordinary substance, say water, contains 
of the order of Avogadro's number of molecules. 
Keeping track of the motion of these many molecules
in the way portrayed by Laplace, namely by tracing in time the
$6N$ variables ${\xx}_i(t)$ and ${\vv}_i(t)$
turns out to be a scientific {\it chimera}.
Even assuming one has enough capacity to store so much information,
one would still be left with the problem of dynamic instabilities
in phase space: any tiny uncertainty in the initial microscopic state
would blow up exponentially in time, thereby shrinking
the predictability horizon of the system virtually to zero.
It is a great gift that such a huge amount of information, besides  
being unmanageable, is also needless as well, as we shall see
in the next section.

\subsection{Many-body kinetic level}

The atomistic level deals with molecular positions and speeds
and is governed by the Newton-Hamilton equations 
which describe a world of {\it trajectories}.
The $N$-body kinetic level deals with {\it distribution functions}
$f_N({\xx}_1,{\vv}_1 \dots {\xx}_N,{\vv}_N,t)$, namely smooth
fields describing the joint probability
to find molecule $1$ at position ${\xx}_1$ with speed ${\vv}_1$,
{\it and} molecule $2$ at position $x_2$ with speed ${\vv}_2$,
and so on up to molecule $N$ around position ${\xx}_N$ 
with speed ${\vv}_N$, all at the same time $t$.
Trajectories are here replaced by the notion of {\it phase-space}
fluids obeying a $6N$ dimensional continuity equation, known
as the {\it Liouville equation}:
\begin{equation}
\label{LIOUVI}
[\partial_t + \sum_{i=1}^N {\vv}_i \cdot \partial_{{\xx}_i} + 
{\ac}_i \cdot \partial_{{\vv}_i}] f_N=0
\end{equation}
where ${\ac}_i={\FF}_i/m_i$ are the molecular accelerations.
The underlying assumption is {\it ergodicity}: the time spent
by the trajectory of the $6N$ dimensional coordinate
$P(t) \equiv [{\xx}_1(t) \dots {\vv}_N(t)]$ in a given 
differential volume element $\Delta \Gamma$ of
phase space, is proportional to the measure of $\Delta \Gamma$.

The Liouville equation does not {\it by any means} reduce
the amount of information to be handled via the Newtonian
approach. In fact, since $f_N$ is a continuum $6N$ dimensional field,
the amount of {\it computational} information blows up exponentially!
The Liouville equation represents nonetheless a very valuable step, 
not because we can solve it, but because it sets the stage for
a very elegant and powerful procedure to
consistently eliminate irrelevant information.
Simply integrate $f_{N}$ over unwanted single-particle
coordinates, to define low-order reduced distribution functions
$f_M \equiv f_{12...M<N}=\int f_{12...N} dz_{M+1}...dz_N$,
where $dz_k \equiv dx_k dv_k$, $k=M+1, \dots N$.
The result is a chain of equations:
\begin{equation}
\label{bbgky}
[\partial_t + \sum_{i=1}^M \vv_i \cdot \partial_{\xx_i} + 
{\ac}_i \cdot \partial_{\vv_i}] f_M = C_M
\end{equation}
known as the BBGKY hierarchy.
Note that the right-hand side collects the effects of 
intermolecular interactions.
In the presence of a $b$-body potential, $C_M$ only involves $b$ upper-lying
distributions $f_{M+1},... f_{M+b}$. 
Fortunately, most interesting macroscopic observables, such as 
density, pressure, temperature, and energy, only depend on $1$ or $2$-body 
distributions, so that our efforts can be channeled into
the bottom floors, $M=1,2$, of the BBGKY hierarchy.

\subsection{The Boltzmann equation and the $H$-theorem}

The most important one-body equation 
is the celebrated Boltzmann equation: 
\begin{equation}
\label{BOL}
\partial_t f + \vv\cdot\partial_{\xx} f + {\ac} \cdot \partial_{\vv} f= C[f,f]
\end{equation}
where $f(\xx,\vv,t)$ is the probability density of finding a classical
pointlike particle at position $\xx$ at time $t$ with speed $\vv$.
The left hand side represents free streaming in phase space $(\xx,\vv)$
and the right hand side denotes the effects of binary collisions, typically
a very complicated integral operator encoding the details of molecular
interactions. 
The Boltzmann equation relies on
the famous molecular chaos ({\it ``Stosszahlansatz''})
assumption:
\begin{equation}
f_{12}(\xx_1,\vv_1,\xx_2,\vv_2,t)=f(\xx_1,\vv_1,t) f(\xx_2,\vv_2,t)
\end{equation}
which asserts the absence of correlations between molecules
entering a binary collision.
It is precisely this arbitrary--if plausible--
assumption which breaks time-reversal symmetry, since it is clear
that {\it after} a collision molecules must be correlated 
because of mass-momentum-energy conservation.
%SSSSSSSSSSSSSSSSSSSSSSSSSS
The essence of the molecular chaos assumption is that 
these post-collisional correlations decay exponentially fast in time
so that the probability of the two particles to collide with
each other again in a correlated state
after any finite time lapse is virtually zero.
%SSSSSSSSSSSSSSSSSSSSSSSSSS
Breaking time-reversal symmetry opens
the door to irreversible behavior, and one of the
most profound of Boltzmann's contributions to
statistical mechanics rests with his discovery of a quantitative measure
of irreversibility: the celebrated $H$-theorem.
This quantitative measure of irreversibility is provided by the
Boltzmann $H$-function (in what follows, we term it also as 
the entropy function; the physical entropy being $S=-k_{\rm B}H$):
\begin{equation}
H(t) = \int f(\xx,\vv,t) \ln f(\xx,\vv,t) d\vv d\xx
\end{equation}
which was shown by Boltzmann to be monotonically non-increasing
in time, $dH/dt \leq 0$, {\it regardless} of the details of
the collision operator. 
The $H$-theorem stands out as a conceptual
bridge between micro- and macro-dynamics. 
Yet, it is difficult to think of a more debated 
and controversial issue in theoretical physics 
%SSSSSSSSSSSSSSSSSSSSSSSSSS
(see Wehrl, 1978).
%SSSSSSSSSSSSSSSSSSSSSSSSSS
We will not delve here into the details of the various 
paradoxes which were raised against 
the $H$-theorem, nor shall we discuss the fact 
that Boltzmann derived his theorem without 
demonstrating under which conditions his equation, 
a complicated integro-differential initial-value 
problem, does indeed have solutions. 
While leaving mathematical rigor somehow behind, the      
$H$-theorem is nonetheless a monumental
contribution to modern science, since it 
showed for the first time the way to
a grand-unification of two fundamental and hitherto disconnected 
domains of science: Mechanics and Thermodynamics. 

The practical importance of Boltzmann-like equations was
only to be furthered by modern developments in theoretical
physics, primarily the emergence of the fundamental
notion of {\it quasi-particles} as collective excitations
of non-linear field theories (Kadanoff and Baym, 1962).
By shifting the focus from actual particles (real atoms or molecules)
to quasi-particles, the Boltzmann equation goes way beyond the original 
framework from which it was derived 
(i.e., rarefied gas dynamics) and spreads nowadays 
its wings across a huge variety of fields in statistical
mechanics, including neutron and radiation transport, electron 
transport in semiconductors, hadronic plasmas, and many others.
Quasi-particles play a central role also 
in the top-down approach to statistical mechanics 
that is to be described shortly.

\subsection{The macroscopic level}

Macroscopic observables such as fluid mass density, 
speed and energy density are obtained from the one-body
kinetic distribution by integration over velocity space:
\begin{eqnarray}
\label{nue}
\rho(\xx,t) = m \int f(\xx,\vv,t) d\vv,\\\nonumber
\rho \uu(\xx,t)= m \int f(\xx,\vv,t)\vv d\vv,\\\nonumber
\rho e(\xx,t)=m \int f(\xx,\vv,t) \frac{v^2}{2} d\vv
\end{eqnarray}
where $m$ is the atomic/molecular mass.
Supplementing these formal integrations with 
additional assumptions (e.g., small deviations from local
thermodynamic equilibrium), one finally arrives at the desired
equations for the macroscopic observables, typically 
the Navier Stokes equations of fluid dynamics (for the time being,
we restrict ourselves to the case of isothermal fluids, for which
the energy equation is not needed):
\begin{eqnarray}
\label{NSE1}
\partial_t \rho + {\rm div}(\rho {\uu})=0,\\\nonumber
\partial_t \rho {\uu} + {\rm div}(\rho {\uu} {\uu})
=-\nabla P + {\rm div}(2\mu\overline{\nabla {\uu}}
+ \lambda \UNIT {\rm div}{\uu}),
\end{eqnarray}
where $\rho$ is the fluid density, ${\uu}$ the fluid speed,
$P$ the fluid pressure, the overline denotes the symmetric 
dyad $\nabla {\uu} + (\nabla {\uu})^T$, the superscript $T$ 
indicates the transpose, $\mu$ and $\lambda$ are
the shear and bulk dynamic viscosities respectively, 
and $\UNIT$ denotes the tensor identity.
The Navier-Stokes equations keep no track of the
discrete nature of the underlying microscopic world, and 
are paramount for the quantitative description of macroscopic systems.

\subsection{The top-down approach}

Each jump down the BBGKY ladder 
removes irrelevant information so that, in the end, the
$10^{23}$ atomistic trajectories are replaced by 
the evolution of a handful of continuum hydrodynamic fields.

The BBGKY approach is formally elegant and 
very fruitful for further theoretical insight and analysis.
In fact, it is perfectly positioned to borrow the powerful
mathematical machinery of classical and
quantum statistical field theory, such as perturbative 
methods, diagrammatic techniques and the like.
Less noted, perhaps, is the fact that the 
resulting equations prove {\it exceedingly difficult}
to solve in actual practice. 
This is true even at the coarsest level: 
the Navier-Stokes equations are notorious for posing 
one of the hardest problems left in classical 
(i.e., non quantum) physics, namely fluid turbulence.
It makes sense therefore to think of complementary
routes to BBGKY, more aligned with the spirit of model-building and
computational tractability rather than with amenability to analytical treatment.

An emerging and still fast-developing strategy along these lines
is provided by {\it fictitious dynamics} methods.
The idea is to introduce effective molecules
({\it ``computational quasi-particles''}), each representing 
a huge number, say $R$, of real ones, so that the number of effective 
molecules we must deal with is no longer of the order
of Avogadro's number, but of order $N_R=N_A/R << N_A$ instead.
In terms of these effective molecules, the Newtonian equations 
are as follows:
\begin{equation}
M_I \frac{d^2 \XX_I}{d t^2} = {\FF}_I (\XX) 
+ \DD_I,\;\;\;I=1,N_R
\end{equation}
where $\XX_I$ represents a coarse-grained coordinate, 
and $M_I=\sum_{i=1}^R m_i$ is the total mass of the
effective ``macro-molecule''.
The term $\DD_I$ collects all the details of the underlying 
fine-scales and disappears only in the trivial case of a linear
dependence of the force ${\FF}_I$ on the molecular coordinates.
Realistic forces are generally inverse powers of the intermolecular
distance and, consequently, $\DD_I \ne 0$.
A proper renormalization/closure procedure would attempt to incorporate the
effects of the fine scales into appropriate (and most likely very
complicated) expressions for a renormalized force 
$\tilde {\FF}_I = {\FF}_I + \hat C {\DD}_I$, $\hat C$ 
being some form of projection operator making the renormalized force
available in terms of the coarse-grained coordinates. 
Most simply, one sets ${\DD}_I=0$ and proceeds with the solution
of the {\it same} Newtonian equations, only applied to a much smaller
set of molecules: this is the very successful path
taken by Molecular Dynamics (Alder and Wainwright, 1959).

However, once it is agreed that the ultimate aim is
macroscopic physics, say solving the Navier-Stokes equations, even
Molecular Dynamics is still redundant of needless micro-details.
We may want more than just setting ${\DD}_I=0$, and look 
for instance for expressions of the coarse-grained
force $\tilde {\FF}_I$ that are simpler than those associated
with true intermolecular interactions.
The principle of Occam's razor, i.e., maximum simplicity
implies that one should choose the 
simplest coarse-grained dynamics 
compatible with the target macroscopic equations.
This is the hard-core idea of ``minimal molecular dynamics'':
relinquish as many microscopic details as possible right
at the outset (atomistic level), making sure, however,
that the basic symmetries, conservation laws and
evolutionary constraints needed to ensure the correct 
macroscopic behaviour are preserved in the process.
As to the first principle of thermodynamics, this means 
conserving all-and only-the microscopic invariants 
(mass, momentum, energy), while 
compliance with the second principle of thermodynamics
imposes the existence of a
suitable monotonically increasing function of time,
the Boltzmann $H$-function and the related entropy.
The actual realization of the top-down approach
is by no means unique (for a recent form of
dissipative particle dynamics, see Hoogerbrugge and
Koelman, 1992, Espanol and Warren, 1995,
Flekkoy and Coveney, 1999).
In the following, however, we shall refer to 
Lattice Gas Cellular Automata (LGCA), a particularly appealing 
instance of minimal particle dynamics developed in the mid 1980's, which 
provided the roots of the Lattice Boltzmann (LB) method.

\section{Lattice Gas Cellular Automata}

The theory of lattice Gas Cellular Automata is a 
rich subject and has been recently made accessible in full
detail by a series of beautiful monographs 
(Rothman and Zaleski, 1997, Chopard and Droz, 1998, 
Rivet and Boon, 2001).
Here we shall take a substantial shortcut and proceed by example.

\subsection{The Frisch-Hasslacher-Pomeau automaton}

Let us begin by considering a regular lattice 
with hexagonal symmetry such that each
lattice site is surrounded by six neighbors 
identified by six connecting
speeds ${\vv}_i \equiv v_{ia},\;i=1,6$,
the index $a=1,2$ running over the spatial dimensions $x,y$
(see Figure 2, which also includes a rest-particle with
zero speed).
%\begin{figure}
%\includegraphics{f2_1.eps}
%\caption{The FHP hexagonal lattice.}
%\end{figure}
Each lattice site hosts up to six particles with the following
prescriptions
\begin{itemize}
\item All particles have the same mass $m=1$
\item Particles can move only along one of the six directions
      defined by the discrete displacements $v_{ia}$.
\item In a time-cycle (set at unity for convenience) the particles hop
      synchronously to the nearest neighbor in the
      direction of the corresponding discrete vector $v_{ia}$. 
      Longer or shorter jumps are both forbidden, which
      means all lattice particles have the same energy. 
\item No two particles sitting on the same 
      site can move along the same direction $v_{ia}$
      ({\it Exclusion Principle}).
\end{itemize}
These prescriptions identify a very stylized 
gas analogue, whose dynamics is made purposely unaware
of microscopic details of real-molecule Newtonian dynamics.
In a real gas molecules move along any direction 
({\it isotropy}) whereas here they are confined to 
a hexagonal cage. Also, real molecules can move
virtually at any (subluminal) speed, whereas here 
only six mono-energetic beams are allowed.
Amazingly, this apparently poor representation
of true molecular dynamics has all it takes to
simulate realistic hydrodynamics!
With the prescription kit given above, the state 
of the system at each lattice site is unambiguously 
specified in terms of a plain {\em yes/no} option saying
whether or not a particle sits on the given site.
This dichotomic situation is readily
coded with a single binary-digit (bit) per site and direction
so that the entire state of the lattice gas is specified by
$6N$ bits, $N$ being the number of lattice sites.
Borrowing the parlance of second quantization, we 
introduce an {\it occupation number} $n_i$, such that
\begin{eqnarray}
n_i ({\xx},t)= \lbrace 1,0 \rbrace
\end{eqnarray}
depending on whether or not the
lattice site $\xx$ hosts a particle with speed $\vv_i$ at time $t$.
The collection of occupation numbers
$n_i({\xx},t)$ over the entire
lattice defines a $6N$ dimensional time-dependent
boolean field whose evolution takes place in a boolean
phase-space consisting of $2^{6N}$ discrete states. 
This boolean field takes the intriguing
name of {\it Cellular Automaton} 
to emphasize the idea that not only space and 
time, but also the dependent variables (matter)
take on discrete (boolean) values.
The fine-grain microdynamics of this boolean field {\it cannot}
be expected to reproduce the true molecular dynamics to any
reasonable degree of microscopic accuracy.
However, as is known since Gibbs, many different microscopic
systems can give rise to the same macroscopic dynamics, and
it can therefore be hoped that the macroscopic dynamics 
of the lattice boolean field would replicate real-life 
hydrodynamic motion even if its microdynamics does not.

\subsection{Boolean microdynamics}

Let us now prescribe the evolution rules of our
cellular automaton. Since we aim at hydrodynamics, we 
should address two basic mechanisms: 

- {\it Free-streaming}

- {\it Collisions}

Free streaming consists of simple particle transfers from site
to site with discrete speeds ${\vv}_i$.
Thus, a particle sitting at site ${\xx}$ at time $t$ with
speed $v_{ia}$ will move to site 
${\xx}+{\vv}_i$ at time $t+1$.

In equations:
\begin{equation}
\label{STREAM}
n_i ({\xx}+{\vv}_i,t+1) = n_i({\xx},t)
\end{equation}
This defines the discrete free-streaming operator
$S_i$ as
\begin{equation}
\label{STREAM2}
S_i n_i \equiv n_i ({\xx}+{\vv}_i,t+1) - n_i({\xx},t)
\end{equation}
This operator is a direct transcription of the 
Boltzmann free-streaming operator, 
$D_t \equiv \partial_t + v_a \partial_a$,
to a discrete lattice in which space-time variables are discretized
according to the synchronous ``light-cone'' rule: 

\begin{equation}
\label{SYNCH}
\Delta x_{ia}=v_{ia} \Delta t.
\end{equation}
Once on the same site, particles interact
and reshuffle their momenta so as to exchange mass and momentum 
among the different directions allowed by the lattice
(see Figure 3).

%\begin{figure}
%\includegraphics{f2_4.eps}
%\caption{A FHP collision with two equivalent outcomes.}
%\end{figure}

This mimics the real-life collisions 
taking place in a real gas, with the
crude restriction that all pre- and post-collisional momenta 
are forced to '`live'' on the lattice.
As compared with continuum kinetic theory, 
Lattice Gas Cellular Automata introduces a very radical cut of degrees of 
freedom
in momentum space: just one speed 
(all discrete speeds share the same
magnitude $c=1$, hence the same energy) 
and only six different propagation angles.
Not bad for an original set of six-fold infinite
degrees of freedom!
Space-time is also discretized (see (\ref{SYNCH})), but 
this is common to all computer simulations of dynamical systems.
At this stage, it is still hard to believe that such a stylized 
system can display all of the complexities
of fluid phenomena. And yet it does!
The reader acquainted with modern statistical mechanics
smells the sweet scent of {\it universality}: for all
its simplicity, the Frisch-Hasslacher-Pomeau
automaton may display the same large-scale properties
of a real fluid (Kadanoff, 1986), such as propagation
of sound waves, vortex interactions and energy dissipation.
The name of this magic is {\it symmetry and conservation}.

Let us dig a bit deeper into this matter.

Let us consider the Frisch-Hasslacher-Pomeau
collision depicted in Figure 3:
Albeit stylized, this collision shares two crucial features 
with a real molecular collision:
\begin{itemize}
\item It conserves particle number (2 before, 2 after)
\item It conserves total momentum (0 before, 0 after)
\end{itemize}
Symbolically, its effect on the occupation 
numbers is a change from $n_i$ to $n'_i$ on the same site
\begin{equation}
\label{COLLI}
n'_i -n_i = C_i (\underline n)
\end{equation}
where $\underline n \equiv [n_1,n_2 \dots n_b]$ denotes
the set of occupation numbers at a given lattice site.

To sum up, the final Lattice Gas Cellular Automata 
update rule reads as follows:

\begin{equation}
\label{LLIOUV}
S_i n_i=C_i
\end{equation}
or, which is the same:
\begin{equation}
\label{LLIOUV2}
n_i ({\xx}+{\vv}_i,t+1)=n'_i ({\xx},t)
\end{equation}
where all quantities have been defined previously.
The equations (\ref{LLIOUV}, \ref{LLIOUV2}) represent the 
microdynamic equation for the boolean lattice gas, the analogue 
of Newton equations for real molecules.
This equation constitutes the starting point 
of a Lattice BBGKY hierarchy,
ending up with the Navier-Stokes equations. 
At each level, one formulates a lattice counterpart of the 
various approximations pertaining to the four levels 
of the hierarchy (see Figure 1).

\subsection{Merits and pitfalls of Lattice Gas Automata}

The major appeals of Lattice Gas Cellular Automata are:
{\it
\begin{itemize}
\item Round-off free computing (boolean algebra is exact)
\item Memory savings (only one bit per degree of freedom)
\item Virtually unlimited potential for parallel computing 
\end{itemize}
}
These points fueled the excitment around Lattice
Gas Cellular Automata as a potentially
revolutionary tool for computational fluid dynamics (and more).
In particular, the hope was to attack the infamous problem
of fluid turbulence. Let us just mention that 
this relates to the dynamics of flows where dissipative effects are
very small as compared with advection.
The relative strength of advection versus dissipation is measured by a 
dimensionless number known as the Reynolds number:
\begin{equation}
Re = \frac{UL}{\nu}
\end{equation}
where $U$ is a typical flow speed on a macroscopic scale $L$ (the size
of the device) and $\nu$ is the kinematic viscosity of the fluid.
Based on dimensional scaling theories 
(A.N. Kolmogorov, 1941), it can be shown
that the number of degrees of freedom associated with a turbulent
flow at a given Reynolds number $Re$ is approximately $Re^{9/4}$.
Since $Re \sim 10^6$ is a commonplace in daily life (that's 
more or less
what we experience by driving our car at cruising speed of about $100$
Km/h in full compliance with traffic regulations...), the simulation
of such flows implies the solution of about $10^{14}$ degrees of freedom:
less than Avogadro's number, to be sure, but still too 
much for any foreseeable computer.
These figures say it all as to the hunger for innovative mathematical
and numerical methods for fluid turbulence!
Unfortunately, closer inspection into the details
of the Lattice Gas Cellular Automata
method reveal a number of difficult problems:
{\em
\begin{itemize}
\item Statistical noise
\item Complexity of the collision operator
\item Small number of collisions
\end{itemize}
}

Statistical noise relates to the fact that in order to extract
a smooth hydrodynamic signal, averages over {\it many} boolean
variables are required, thereby eclipsing the memory savings provided
by the boolean microdynamic representation. 

Exponential complexity relates to the exponential escalation of
the collision operator as more physics is added to the model,
say, more than one fluid species, or simply by moving to higher 
dimensions. 
The problem is that the complexity of the corresponding collision operator
grows exponentially, roughly as $2^b$, where $b$ is the number of bits
per site, so that the original simplicity is rapidly lost.

Low collisionality is also related to the paucity of discrete speeds.
Only a relatively small fraction of phase space is
collisionally active since many collisions are simply not compatible
with conservation principles. Few collisions mean long mean-free-path,
hence high momentum diffusivity $\nu$, hence low-Reynolds numbers,
and the dream of simulating highly turbulent flows fades away.

In spite of the remarkable progresses achieved in the
late 1980's, the Lattice Gas Cellular Automata algorithms 
remained rather heavy and stiff with respect to the collision rules. 
On the other hand, these inconveniences were not compensated
by a dramatic advantage in terms of accessible Reynolds
numbers. As a result, the interest in Lattice
Gas Cellular Automata as a tool for high 
Reynolds flow simulations leveled off in the early 1990's. 
This was precisely the time the Lattice Boltzmann method took off.

\section{Lattice Boltzmann equations}

The theory of the lattice Boltzmann equation (LBE) 
(for a review see, Benzi, Succi, Vegassola, 1992,
Qian, Succi, Orszag, 1995 and Chen, Doolen, 1998) begins with
a straightforward floating point recast of the boolean
evolution equation of Lattice Gas Cellular Automata dynamics 
(Frisch, Hasslacher and Pomeau, 1986; 
Frisch {\em et al.}, 1987).
With reference to a set of $b$ populations, $f_i({\xx},t)$, with
$\{i=1, \ldots , b\}$ representing the probability
for a particle to reside on a lattice site ${\xx}$
at time $t$ with discrete velocity ${\vv}_i$, the LBE reads
as follows:
\begin{equation}
\label{LBEnl}
f_i({\xx} + {\vv}_i, t + 1) = f_i({\xx},t)
+ C_i[f_1, \ldots , f_b],
% \;\;\; i= 1, \ldots , b
\end{equation}
where $C_i[f]$ is the collision operator,
a polynomial of degree $b$ constructed explicitly out
of all allowable $n$-body collisions, $2 \leq n \leq b$,
among populations
sitting on the {\it same} site ${\xx}$ at time $t$.
The above LBE has a simple interpretation: Particle
population $f_i({\xx}+{\\v}_i, t+1)$ is equal to the 
post-collision $f'_i ({\xx}, t)$ value
advected from the ``upper wind'' position ${\xx}$
at the previous time $t$. Here $f'_i \equiv f_i + C_i$.
Thus $C_i$ represents the change of the particle
population by collisions.
Compliance with mass and momentum conservation laws imposes the
following constraints on the collision operator:
\begin{equation}
\label{CON}
\sum_i C_i = 0,\; \sum_i {\vv}_i C_i = 0
\end{equation}
If there is an energy degree of freedom, then
we also have:
\begin{equation}
\sum_i \epsilon_i C_i = 0,
\end{equation} where $\epsilon_i = {\vv}_i^2/2$ for an
ideal gas.
As in continuum kinetic theory, this collision
operator admits a detailed balance condition
when all populations of different particle velocities
are in equilibrium, $\{ f_i = f^{\rm eq}_i, \;\; \forall i\}$:
\begin{equation}
\label{Eql}
C_i[f^{\rm eq}] = 0
\end{equation}
Regardless of the detailed collision processes,
the solution of Eq.\ (\ref{Eql}) takes on a generic form
dictated by the basic conservation laws,
\begin{equation}
\label{LEQnl}
f_i^{\rm eq} = e^{-I_i}
\end{equation}
where $I_i=A+{\bf B} \cdot {\vv}_i + C\epsilon_i$ is a
linear combination of collisional invariants, while  $A$ and
${\bf B}$, in turn, are functions of local hydrodynamic quantities,
\begin{equation}
\rho ({\xx}, t) = \sum_i f_i ({\xx}, t)
\end{equation}
\begin{equation}
\rho \uu ({\xx}, t) = \sum_i {\vv}_i f_i ({\xx}, t)
\end{equation}
Once again, for systems with an energy degree of freedom,
the total energy can be defined as,
\begin{equation}
\rho (D T + \uu^2)/2 = \sum_i \epsilon_i f_i
\end{equation}
where $D$ is the number of spatial dimensions
and $T$ can be interpreted as the temperature of the fluid.

The LB equation (\ref{LBEnl}) 
obeys the so-called semi-detailed balance property for the collision operator
as its Lattice Gas Cellular Automata predecessor 
(Frisch, Hasslacher and Pomeau, 1986; 
Frisch {\em et al.}, 1987). 
It can be shown that such a kind of 
collision operator admits a local $H$ theorem
with the following discrete Boltzmann entropy function,
$h_{\rm B}({\xx},t) = \sum_i f_i({\xx}, t) \ln f_i({\xx}, t)$.
In other words, an $H$ theorem can be defined on each
local lattice site, so that, without external
disturbances or boundary influences, $h_{\rm B}$ is a non-increasing
function of $t$ satisfying the dynamics of $C_i$.
Furthermore, the minimum value of $h_{\rm B}(\xx,t)$ is attained
when the particle populations $\{ f_i\}$ assume the equilibrium
form given by (\ref{LEQnl}), where $A$, ${\bf B}$ and $C$ are
constants determined by the values of mass, momentum
and energy. Borrowing from the terminology
of nonlinear dynamic systems, the existence
of an $H$ theorem ensures that the equilibrium
distribution is not only a {\it fixed point} solution of (\ref{Eql})
but also an {\it attractor} of the collisional dynamics. 
On the other hand, this is a weaker
$H$ theorem compared to its continuum counterpart, for
the above does not include advection of particles among lattice sites
according to (\ref{LBEnl}). We will realize in a subsequent
section that, unlike in continuum Boltzmann dynamics,
the local $H$ theorem in Lattice Gas Cellular Automata 
given above does {\it not} (except for the
isothermal case, in which energy conservation is replaced
by a constant temperature-like parameter) automatically lead to
a global one, referred to as the global $H$-theorem 
(Frisch, Hasslacher and Pomeau, 1986; 
Frisch {\em et al.}, 1987,Chen, 1995; Chen, 1997).
Generally the global $H$ function is given by,
\[ H(t) = \sum_{\xx} h_{\rm B}({\xx},t). \]
Existence of a global $H$ theorem in LB
guarantees an asymptotically homogeneous spatial
distribution of particles as time $t \rightarrow \infty$,
and hence it provides a well-defined global hydrodynamic stability.
As in standard kinetic theory, hydrodynamics concerns
properties around the local equilibrium,
\begin{equation}
f_i = f_i^{\rm eq} + f_i^{\rm ne}
\end{equation}
where the non-equilibrium component, $f_i^{\rm ne}$,
is supposed to scale like $k f_i^{\rm eq}$, where the smallness
parameter $k$, known as the Knudsen number, is the ratio 
of the particle mean free
path to a typical hydrodynamic scale. Ordinary fluids feature
Knudsen numbers below $0.01$, see Chapman and Cowling, 1970.) 
The existence of a local
$H$ theorem makes a perturbation analysis purposeful since
the dynamic system always approaches a defined equilibrium
$f_i^{\rm eq}$ and evolves in the neighborhood of such a defined point.
Perturbative treatment of the LBE up to second order in $k$
(Chapman-Enskog procedure)
is then expected to yield the hydrodynamic equations
\begin{eqnarray}
\label{NSE}
\partial_t \rho + \partial_a \rho u_a = 0 \nonumber \\
\partial_t \rho u_a + \partial_b P_{ab} = 0
\end{eqnarray}
where
\begin{eqnarray}
P_{ab}&\equiv&\sum_i v_{ia} v_{ib} f_i^{\rm eq}
+ \sum_i v_{ia} v_{ib} f_i^{\rm ne}\\\nonumber 
&=&g\rho u_a u_b + P \delta_{ab} 
+ \rho \nu [\partial_a u_b+\partial_b u_a
- \frac{1}{D} (\partial_c u_c)  \delta_{ab}] 
\end{eqnarray}
is the momentum-flux tensor comprising both non-viscous and dissipative
components, the latter being proportional to
the fluid viscosity $\nu$. In the above, $P$ is
the resulting pressure, and $D$ is the dimensionality
of a lattice. Note that the prefactor $g$ in the advection term
which signals a potential breaking of Galilean invariance 
whenever $g \ne 1$. On the other hand, since
$g$ depends only on the fluid density $\rho$, for incompressible
flows in which the density is constant in space and time, Galilean 
invariance can be recovered by a simple rescaling of time, pressure 
and viscosity:
\[
t'= gt,\;\;\;\nu'=\nu/g,\;\;\;P'=P/g
\]
Clearly, even within the framework of incompressible
flows, this rescaling does {\it not} extend to more general
situations, such as multi-component or multi-phase flows.

\subsection{The fully non-linear LBE}

The fully non-linear LBE, eq. (\ref{LBEnl}), was first proposed
by McNamara and Zanetti (McNamara, Zanetti, 1988), who realized
its potential for doing away with the statistical noise problem
affecting Lattice Gas Cellular Automata simulations (Orszag, Yakhot, 1986, 
Succi, Santangelo, Benzi, 1989).  
All other weaknesses still remained.
In general, the factor $g$ is not equal to unity, indicating a violation of
Galilean invariance. Though a choice of sufficient lattice symmetry
ensures a rotationally invariant
form of $P_{ab}$, the resulting local
equilibrium in typical Lattice Gas Cellular Automata
collisions does not have
a suitable functional form for achieving a
correct Navier-Stokes hydrodynamics, which requires $g=1$.
Indeed, in order to obtain the correct form of the
Navier-Stokes equation, the
local equilibria must comply with a specific form
as a function of the local hydrodynamic variables, $\rho$
and $\uu$, (S.\ Chen {\em et al.}, 1991;
H.\ Chen, S.\ Chen and Matthaeus, 1992; 
Qian, d'Humieres and Lallemand, 1992).

In addition to incorrect hydrodynamics, Lattice Gas
collisions have problems in attaining low viscosity.
The fluid viscosity is approximately given by
$\nu =3D l v_T$, where $l$ is the particle mean-free path and
$v_T$ the typical particle thermal speed.
In many problems, for instance turbulence, we are interested in 
very low-viscous flows, which means very short mean-free paths. 
On the other hand, short mean-free paths imply many
collisions.

We are thus forced to consider as far as possible all allowable
types of collisions among the set of discrete velocities.
Unfortunately, it turns out that the full collision operator $C_i$
involves an exponential barrier of the order of $2^b$ operations, in
direct contradiction with the basic commitment to
simplicity and computational efficiency of the method.

\subsection{The LBE in scattering form}

Building upon the idea that many-body collisions are not essential
to achieve the correct hydrodynamic limit,
Higuera and Jimenez (Higuera and Jimenez, 1989) realized that 
the collision operator can be reduced to a dramatically 
simpler two-body scattering expression:
\begin{equation}
C_i \rightarrow  \sum_{j}A_{ij} [f_j-f_j^{\rm eq}],
\end{equation}
where the scattering matrix $A_{ij}$ is basically the Jacobian of the
fully non-linear operator $C_i$ evaluated at the uniform equilibrium
values $f_i=\rho/b$.
The above expression turns a daunting $2^b$ complexity into a much
more manageable $b^2$ one, thus opening the way to three-dimensional
Lattice Boltzmann hydrodynamics.
Of course, compliance with an $H$-theorem is no longer guaranteed
because the local equilibrium is no longer the direct result
of collisional dynamics.

In addition, since the scattering matrix $A_{ij}$ is still 
related one-to-one
to the underlying Lattice Gas Cellular Automata microdynamics, the corresponding 
Lattice Boltzmann equation shares
the same limitations in terms of high viscosity, i.e., 
low Reynolds numbers. 

\subsection{The self-consistent LBE}

This last limitation can be wiped out by a mere change in perspective.
Instead of deriving the Navier-Stokes equation
bottom-up (here bottom means the atomic level) from a truly
$N$-body discrete dynamical system, we can {\it construct}
it top-down from the sole requirement of compliance with
the Navier-Stokes equations
(Higuera, Succi and Benzi, 1989, Succi, Benzi, Higuera, 1991).
The idea is to recognize that, as far as
hydrodynamics is concerned, the key notions of scattering
matrix and local equilibrium can be {\it prescribed} at 
the outset instead of being {\it
derived} from an underlying (discrete) micro-dynamics. 
Mathematically, this amounts to prescribing the scattering
matrix in spectral-form:
\begin{equation}
A_{ij}=\sum_{k=1}^b \lambda_k P_{ij}^{(k)},
\end{equation}
where $P_{ij}^{(k)}$ projects along the $k$-th eigenvector
in kinetic space and $\lambda_k$ is the 
corresponding eigenvalue ($\lambda_k=0$ for conserved quantities).
Of particular interest is the leading non-zero eigenvalue, which
is in direct control of the fluid viscosity $\nu \sim \lambda^{-1}$.

Once this point of view is endorsed, namely that local equilibria can be
``freely" chosen within the conservation constraints, and that
the scattering matrix can be selected a priori on the sole
basis of conservation constraints, nothing prevents this
Lattice Boltzmann equation from attaining as low viscosity as the lattice 
discreteness permits.
Full contact with classical computational 
fluid dynamics is made (Chen H., S. Chen, W. Matthaeus, 1992).
More importantly, the top-down approach, which proved so fruitful
and influential for subsequent developments of LB theory, is established. 

\subsection{The lattice Bhatnagar-Gross-Krook equation}

The LBE story has still one more nugget in store.

The scattering matrix can also be viewed as a multiple scale relaxation
operator, one scale for each non-zero eigenvalue.
Since we are basically interested in a single transport parameter, the
fluid viscosity, a single eigenvalue should do.
Indeed we can replace the full matrix $A_{ij}$ with a single-parameter
diagonal form $-\omega \delta_{ij}$, describing a single-time
relaxation around a prescribed local
equilibrium $f_i^{\rm eq}$. In its simplest and by now most popular form,
the relaxation-approximation corresponds to the following Lattice
BGK (Bhatnagar-Gross-Krook) equation
(Bhatnagar, Gross and Krook, 1954):
\begin{equation}
\label{LBGK}
f_i({\xx} + {\vv}_i,t+1)
- f_i({\xx},t)= -\omega [f_i({\xx},t)-f_i^{\rm eq}({\xx},t)],
%\;\;\;i=1, \ldots , b,
\end{equation}
where exact Navier-Stokes hydrodynamics is obtained
with the following choice of local-equilibrium form,
\begin{equation}
f_i^{\rm eq} = w_i \rho \left\{1 + \frac{{\vv}_i \cdot \uu}{T}+
\frac{({\vv}_i{\vv}_i - v^2):\uu\uu}{2T^2}\right\}.
%\ i = 1, \ldots , b.
\end{equation}
where the symbol $:$ stands for tensor scalar product.
In the above, $w_i$ is a suitable lattice-dependent weighting 
factor, and the temperature $T=1/3$ in typical isothermal lattice BGK models 
(S.\ Chen {\em et al.}, 1991;
H.\ Chen, S.\ Chen and Matthaeus, 1992; 
Qian, d'Humieres and Lallemand, 1992; Chen, Teixeira and Molvig, 1997).
Since temperature is frozen to a constant value, these
lattices models should best be denoted as ``a-thermal''
rather than ``iso-thermal''. 
More comments on this delicate issue shall be presented later in this paper.
For a thorough and beautiful discussion of the subtleties of
lattice thermo-hydrodynamics, the reader is recommended to
consult Rivet \& Boon, Chapter IV. 

It is readily recognized that this formulation leads to a
Galilean invariant Navier-Stokes equation (up to terms of order $Mach^4$,
where $Mach$ is the Mach number, namely the ratio of fluid to sound
speed), for a fluid of viscosity $\nu \sim T/\omega$.
The LBE scheme in relaxation form has met
with significant success in the last decade in simulating
a variety of fluid flows.
Indeed, the lattice BGK equation
and subsequent straightforward extensions
to allow a variable Prandtl number 
(ratio of momentum to heat diffusivity)
(Keizer, 1987, Chen et al 1997), represents the method 
of choice in the field.
Very interesting variants, combining the best features
of the Lattice Boltzmann equation in scattering form 
and Lattice BGK have also been developed (D'Humi\`{e}res, 1992).

A few years later it was shown that the Lattice BGK can
be derived from the continuum Boltzmann BGK equation
by a Grad-like moment expansion supplemented with numerical
quadrature for the actual evaluation of the kinetic moments
(He and Luo, 1997; Shan and He, 1998).
Realizing such a connection might prove very useful 
for establishing new lattice BGK's starting from model continuum
Boltzmann equations, hopefully including physics beyond
the hydrodynamic level.

While focus on the conservation laws
(hydrodynamic constraints) was the leit motif of all
the aforementioned developments, compliance with the second thermodynamic
principle, namely the existence of the $H$ theorem, was
somehow left in a subdued light by the top-down approach.
That neglect had no serious consequences, because, thanks to
a gracious smile from Lady Luck, being cavalier about
an underlying $H$ theorem proved very forgiving 
for {\it isothermal} LBE flows (the bulk of early LBE research) 
even at very low viscosities,
where numerical stability is severely probed.
By now, we have learned that this favorable behavior is due
to the existence of an underlying H theorem, as
we shall detail in the next section.

\section{H theorem in discrete phase-space}

The $H$ theorem is a milestone of non-equilibrium statistical
mechanics, since it provides a conceptual link between the reversible
laws of the microworld and the one-sided nature of macroscopic
phenomena (Lebowitz, 1993; Lieb, 2000). 
It is also a fundamental concept in computational
physics, where compliance with an $H$ theorem is 
often perceived as a byword for numerical stability 
(Perthame and Tadmor, 1991; Natalini, 2000; Junk and Klar, 2000). 
$H$-compliant Discrete-Velocity Boltzmann (DVB) equations
have been known for a long time (Broadwell, 1964, Gatignol, 1965), 
but, as discussed previously, they are not meant
to compete with Navier-Stokes solvers because their main focus is
maximum likelyhood to the Boltzmann equation in the high-Knudsen regime.
This leads to discrete collision operators which are just too complicated
to serve as a practical tool for purely hydrodynamic purposes.

Computational complexity dissolves by using relaxation 
approximations, notably lattice BGK, and it
is therefore imperative to explore the possibility 
of establishing $H$ theorems for lattice BGK. 
Before plunging into this theme, a few
reflections on the classical case are in order.

\subsection{Reflections on the continuum case}

The classical Boltzmann $H$ theorem is so familiar that it is
worthwhile to look at it from the perspective of modeling, in order to
better understand what is actually lost and needs to be rearranged
when going from the continuum to the Lattice Boltzmann case.

First, we find that the familiar local Maxwell
distribution function (the ``Maxwellian'') minimizes 
the $H$ function, $H=\int f\ln f dv$, 
as soon as the five hydrodynamic invariants (mass, momentum
and energy), are fixed.
Furthermore, we observe that all higher-order moments of the local
Maxwellian distribution are just in the right form for the Chapman-Enskog
expansion to deliver the Navier-Stokes equations. 
Backed with this information, we start looking for a kinetic model 
acting as a continuous-time constrained minimization process of this $H$, in 
such
a way as to conserve the hydrodynamic moments. 
We readily find several such models: the Boltzmann equation itself, the BGK
model, the diffusion equation, and so on. All of these models differ
only in the way relaxation to the local
equilibrium takes place; some of them do require explicit knowledge of the
local Maxwellian (the BGK), some others don't. They all deliver the
Navier-Stokes equation in the end, with the only differences
among them occurring in
the transport coefficients, whose evaluation is simple
for the BGK and requires some (non-negligible)
work for the Boltzmann equation.
In all models the collision integral has the local Maxwellian as
its zero point. The rate at which the $H$ function decreases in
time is just equal to the entropy production (we recall that
entropy is the space integral of the $-H$ function)
and entropy production becomes zero in the local Maxwell state. 
The local Maxwellian distribution is then
characterized in three different but equivalent ways: 
{\it It is the minimum of $H$, it is the zero point of 
the collision integral, and it is the zero point of the entropy production}.

When moving to the lattice world the basic question to 
be addressed is: {\it how does the $H$ theorem transform 
in the discrete-time case}?

The first good news is that
{\it all velocities are in some finite range}, so that
we can think of local equilibria for the given density and velocity
alone, without necessarily including energy.
In the (non-relativistic) continuum, energy {\it must}
be included, for otherwise nothing would prevent integrals
over velocities from diverging.
The flip side is that since discrete speeds come in a very(!) 
finite number, the Boltzmann $H$ function does 
{\it not} work: For any known lattice,
a straightforward computation demonstrates that the local
Maxwellian equilibrium does not imply correct expressions for the
higher-order moments. 
On the other hand, a priori, 
there is a huge class of convex functions at our disposal, and
in order to illustrate the idea of what local equilibria may
look like in the lattice context, we shall present an 
example of a solvable lattice entropy,  
$H_{1/2}=\sum_{i=1}^{b}f_i\sqrt{f_i}$ 
\footnote{This entropy function belongs to the class of 
Renyi (or Tsallis) entropies, $H_p=(f^{p+1}-f)/p$, with $p=1/2$.
This is a glorious class of functions of modern statistical
physics, for it provides the stepping stone for the celebrated
replica method in spin-glass theory, multifractal ideas
and also non-extensive thermodynamics (Tsallis, 1988).}.
For this entropy, the local equilibrium can be found explicitly 
(Karlin {\em et al.}, 1998)
\begin{equation}
\label{quadraticFHP}
f_i^{\rm eq}=\frac{\rho}{b}\left[
\frac{1}{2} \left(1+\sqrt{1-M^2}\right)
+ \frac{\uu\cdot{\vv}_{i}}{c_{\rm s}^{2}}+
\frac{(\uu\cdot{\vv}_{i})^2}
{2c_{\rm s}^{4}\left(1+\sqrt{1-M^2}\right)}\right].
\end{equation}
Here $c_{\rm s}^2$ is the speed of sound squared, and
$M^2=u^2/c_{\rm s}^2$ is  the local
Mach number squared.
The above equilibrium is Galilean-invariant only in the limit of
vanishing flow, $M \rightarrow 0$. At any finite flow speed, a
quadratic (in the Mach number) anomaly is apparent. Because
advection terms in the Navier-Stokes equations are also quadratic
in velocity, such a rapid growth of anomalies rules out the use of
the function $H_{1/2}$ for constructing the Lattice
Boltzmann method.  This is the typical situation of
first-generation Lattice Gas and Lattice Boltzman models.
Nevertheless, it is instructive to look at Eq.\
(\ref{quadraticFHP}) and see how lattice equilibria 
differ from the local Maxwellian. 
Recall that the local Maxwellian is a
well-defined function for all values of the average velocity. In 
contrast, the local equilibrium (\ref{quadraticFHP}) is positive
only for $M<1$, and it does not exist as a real-valued
function, for $M>1$. This means that no collision mechanism is
able to equilibrate the non-equilibrium deviations produced by the
supersonic motion, and therefore no macroscopic dynamics 
for this regime can exist. 
This conveys an intriguing flavour of relativistic mechanics,
which is after all not surprising since lattice molecules
move at the lattice speed of light $c$ 
equal to the maximal length of the lattice link.

In order to proceed sensibly, three basic requirements 
must be faced:
{\it
\begin{itemize}
\item Galilean invariance
\item Realizability
\item Solvability
\end{itemize}
}
\noindent each of which we now discuss in some detail.

\subsection{Galilean invariance}

Galilean invariance requires kinetic equilibria to depend on the
relative speed $\vv-\uu$ (``peculiar'' speed)
rather than on the absolute molecular speed $\vv$ itself.
In the continuum, this is ensured by the
Maxwellian dependence,
$\sim \exp\{-(\vv-\uu)^2/v_T^2\}$, where
$v_T \equiv \sqrt{2k_{\rm B}T/m}$ is the
thermal speed, which sets the natural scale for molecular
fluctuations around the fluid speed ${\uu}$.
In the Lattice Boltzmann setting, we have already
committed ourselves to low Mach numbers and no compressibility effects.
Therefore, it seems reasonable to proceed with expansion of the
local Maxwellian around the global equilibrium
(${\uu}=0$). Since the Maxwellian is a transcendental function, 
large departures from global equilibrium require
virtually an infinite number of terms of this expansion.

Terms that correspond to kinetic excitations on top of the uniform
``ground state'' are described by higher order polynomials in the
velocity variable. Since a finite set of discrete speeds can
only support a finite number of these excitations, breaking of
Galilean invariance cannot be avoided. The above considerations
suggest that Galilean invariance can be recouped to some order
if local equilibria can be expressed in the form of finite-order {\it
polynomials} of the flow speed. 

The question is then whether
suitable polynomials can be found that still comply with a
discrete $H$ theorem. 
The answer to this question appears to be negative (Wagner, 1998):
There are no convex entropy functions whose local equilibria are
the polynomials of the form used in the lattice BGK.

In analogy with lattice field theory, we shall
call Galilean-invariant discrete entropies as {\it perfect}, in that they
``hide'' the underlying lattice discreteness. Are there perfect
entropy functions for the Lattice Boltzmann method?
At the time of this writing, no perfect lattice entropy
is known, and most probably there is none, so that we are
led to conclude that the Lattice Boltzmann method 
is not capable of reproducing 
the full properties of the continuum Boltzmann equation. 

However,  
{\it quasi-perfect entropies}, namely entropies which
are not affected by lattice discreteness (up to fourth order terms
in the Mach number) have indeed been found by
a customized, lattice-dependent procedure
(Karlin, Ferrante and \"Ottinger, 1999).
As an illustration, for a three-state one-dimensional
lattice with $\vv_0=0$ and $\vv_{\mp 1} = \mp 1$, the following
quasi-perfect Boltzmann-like $H$-function is identified:
\begin{equation}
\label{result1D3V} 
H= f_{0}\ln (f_{0}/4)+f_{+}\ln f_{+}+f_{-}\ln f_{-}.
\end{equation}
Given the fact that the Lattice Boltzmann equation is itself a second order 
approximation in the Mach number to the Navier-Stokes equations, 
these quasi-perfect entropies must be regarded as definitely  
adequate for hydrodynamic purposes.

The explicit form of the local equilibria corresponding
to the quasi-perfect Boltzmann-like entropy functions is not
known, in general. However, polynomial approximations 
can be found up to the relevant order in Mach number. These
approximations coincide with those established earlier for
the lattice BGK, and this now explains why the Lattice Boltzmann works
in this case: These specific polynomial equilibria survived
among other possible Galilean-compliant polynomials because
they are computationally more stable than others, and
it is precisely these polynomials which are supported by
the quasi-perfect entropy functions. 
The existence of quasi-perfect lattice entropies achieves
a formal compliance with the second law of thermodynamics and
provides a strong incentive
to pursue further development of the Lattice Boltzmann method
based on the entropy maximization principle.

\subsection{Realizability}

Realizability is simply the condition that
local equilibria resulting from an entropy maximization
procedure be real-valued and between zero and one:

\begin{equation}
\label{REA}
0 < f_i^{\rm eq} < 1
\end{equation}

By itself, realizability is neither a {\it necessary} nor a {\it 
sufficient} condition for stability, although it
generally helps stability (Renda et al, 1998).
At any rate, it is a good pre-requisite, at least until we learn to
deal sensibly with negative distribution functions.

It should be appreciated that gradient-driven
departures from local equilibrium can violate the
realizability constraints even if the local equilibria don't.
Therefore the stability domain in kinetic space
is by necessity a subset of the realizability domain.
Manifestly, the polynomial nature of the discrete local
equilibria is a potential danger to the realizability
constraint, a danger which simply does not exist in the continuum.

Description 
of realizability domains on the lattice is a 
nontrivial problem with an interest of its own. 
In a recent work, Boghosian's et. al (2001) applied the powerful
Fourier-Motzkin technique of linear programming 
to classify automatically the realizability
domain associated with a given set of constraints.

This may help the systematic search for optimal lattice entropies.

\subsection{Solvability}

Solvability refers to the possibility of expressing local equilibria
as explicit functions of the conserved
hydrodynamic variables, i.e., density $\rho$ and flow speed
$\uu$. This is very important since it
permits one to explicitly recast kinetic theory in terms of an equivalent
set of partial differential equations for space-time
dependent continuum fields.
Solvability has also a considerable practical impact on the
efficiency of lattice BGK schemes, for it implies that local equilibria
can be encoded once and for all as analytic functions of the
hydrodynamic variables. Local equilibria resulting from
non-solvable entropies must be recomputed numerically at each time
step by iterative procedures and deprive lattice BGK of (part of) its
simplicity and efficiency.
A way out of this problem is to restore the Boltzmann-like
collision operators which can be constructed from just the
knowledge of the entropy. (We recall that even
the single relaxation time property is not exclusively the
property of the Bhatnagar-Gross-Krook equation). 
This has been done recently (Ansumali and Karlin, 2000).

\subsection{Discrete-time effects and the mirage of zero-viscosity}

Perhaps the most interesting feature of the discrete-time kinetics
is that the $H$-theorem is no longer the same (qualitatively) as in
the continuous-time case. The way the $H$-theorem features itself 
in the Discrete Velocity Models, where time is still continous, is 
basically  the same as in the classical theory. 
When time is discrete, things change considerably. 

The major distinction is that, in order to
achieve interestingly low values of the transport coefficients,
the lattice relaxation dynamics must proceed in artificially 
long (hence more effective) jumps going across
the maximum entropy point (equilibrium), in a sort of two-sided
{\it over-relaxation} process, which bears little resemblance
to the smooth relaxational trajectory of the continuum case.
The problem is best illustrated in geometrical terms 
(see Figure 4):
If $f$ is the set of populations at time $t$, then 
the collision operator gives the direction, $C(f)$,  in the 
kinetic space in which the populations must be moved 
(these directions change at different lattice nodes).  
Let us consider populations $f+\beta C[f]$, where $\beta>0$. When tracing 
the ray $f+\beta C[f]$, starting with $\beta=0$, we see that the 
function $H$ is decreasing (because the classical entropy 
production, $-\sum_{i}C_i[f]\partial H/\partial f_i$, is positive),
then comes to the minimum at some $\beta'$, then starts increasing 
which would not be allowed if time were continous. 
In discrete time we can still safely place the populations in 
the increasing branch until $H(f+\beta C[f])<H(f)$), and
finally come to a value $\beta^*$ where the entropy just equals the
initial value $H(f)$. Formally, the condition:

\begin{equation}
\label{entropyest}
H(f)=H(f+\beta^*C[f]),
\end{equation}
sets the limit for the over-relaxation. It has been shown (Karlin et al, 1998;
Karlin, Ferrante and \"Ottinger, 1999; Boghosian et al, 2001) that
this estimate reduces to the so-called linear stability interval, 
$0<\omega<2$ for the lattice BGK,
required by the positivity of transport coefficients, as soon as
the state is close to the local equilibrium.
Recall that in the classical continuous-time kinetic theory,
positivity of transport coefficients is ultimatively related to positivity
of the local entropy production. {\em It is remarkable that the lattice 
{\it local} $H$-theorem establishes the same result 
for the discrete-time case since positivity of transport coefficients
follows now from the estimate (\ref{entropyest}).}
In other words, it is clear that in the over-relaxation 
scenario there is no guarantee that the post-collisional state 
generally attains a smaller $H$-value 
than the pre-collisional state.
This is controlled by the global shape of the entropy 
function {\it far} from equilibrium. 
In fact, if the value of $\omega$ is too large, the end-point
may run into a lower entropy state, or some populations 
may even become negative.
Both such outcomes set the stage for the type of kinetic
instabilities hindering the application of the Lattice Boltzmann method
to very low viscous, high Reynolds flows.

This discussion shows that the ``mirage'' of zero-viscosity,
so crucial for turbulence studies, 
falls within the general problem of finding a systematic
formulation of the dynamics of dissipative
systems {\it far} from equilibrium (Prigogine, 1962; Ruelle, 1999).

With the quasi-perfect entropy at hand, we
make full contact with the second law of thermodynamics, and
establish the ``smallest'' Boltzmann system: It has the properly
defined local equilibrium, the $H$ theorem, and the correct hydrodynamics
to the minimal required order of approximation.

The discrete-time $H$-theorem suggests the possibility of increasing
stability at low viscosity by implementing the estimate (\ref{entropyest}).
In fact, Lattice Boltzmann schemes endowed with quasi-perfect entropies,
and with the entropy estimate (\ref{entropyest}) implemented,
are generally found to exhibit better numerical stability.
Typical instabilities associated with the
lack of the $H$-theorem are shown in Figures 5 and 6. 
In Figure 5 we show the density profile
for a one-dimensional front in a shock tube at
time $t=500$ (in lattice units) for the Lattice Boltzmann
with entropy function (Ansumali and Karlin, 2000) 
(top picture, labelled ELBM),
the Lattice BGK model based on the polynomial ansatz
of Qian, d'Humieres and Lallemand (1992), 
(middle picture labelled LBGK), and the Lattice Boltzmann
model by Qian, D'Humi\`{e}res
and Lallemand, 1991, (bottom figure, labelled LBE).
%SSSSSSSSSSSSSSSSSSSSSSSSSSSSSSSSSSSSSSSSSSSSSSSSS    
The viscosity is set to $\nu=10^{-1}/3$.
From Figure 5, we notice that, although LBE is producing
the most accurate solution almost everywhere, only
ELBM is free of non-physical oscillations
(known as ``Gibbs phenomenon'' in numerical analysis),
typical of non-entropic numerical schemes and often
a precursor of numerical instabilities.
Indeed, a minor decrease of the viscosity is found
to disrupt the stability of the LBE simulation. 
By further decreasing the viscosity,
LBGK also becomes unstable, whereas ELBM does not.
On the other hand, the better ELBM stability comes at
the expense of some oversmoothing of the fronts, as it
is visible by comparison with the thin line (top figure)
giving the exact solution.
The Figure 6, which shows the corresponding velocity profile,
tells essentially the same story: the long-standing conflict
between stability and numerical diffusion, a sort of
``numerical uncertainity principle'' (Boris, 1989).
For very recent improvements of the ELBM approach in the direction
of decoupling the conflicting issues of numerical stability
and accuracy, see Ansumali and Karlin, (2002a, 2002b).
%SSSSSSSSSSSSSSSSSSSSSSSSSSSSSSSSSSSSSSSSSSSSSSSSS    
         
\subsection{Over-constrained equilibria}

We emphasize that local equilibria of the quasi-perfect entropy
are ``classical'': They minimize the entropy under constraints provided
by the locally conserved fields, whereas the desired conditions for
the non-conserved fields (the local equilibrium momentum flux tensor in 
the above consideration) follow from this solution.

{\it What happens if the set of constraints is enlarged 
in such a way as to include higher-order moments?}

It should be noted that this is a rather unconventional move
since the momentum flux tensor does {\it not} originate
from a microscopic collisional invariant.
This move is motivated by the fact that the search
for good entropies cannot keep up with the needs
raised by various Lattice Boltzmann models: while it is rather easy to
establish the set of constraints on the equilibrium, it
is much less easy, even approximately,
to find an entropy whose maximum, under
fixed conserved fields, would also imply the rest of the constraints 
which do {\it not} come from the conservation laws. 
For this reason, one could try to force the solution, proceeding 
with any entropy of choice, but with more constraints.
For instance, it is  clear that minimizers of convex functions under 
constraints, 
\begin{equation}
\label{EXPCON}
\sum_i f^{\rm eq}_i [1,\vv_i,\vv_i\vv_i]= \rho [1,\uu,(P/\rho)\UNIT+\uu\uu],
\end{equation}
are Galilean-invariant {\it by construction} since
they encode the correct equilibrium form of the
momentum-flux tensor right at the outset.
(Any further requirements on the equilibrium can be
added in the same manner.)
The picture is clear:
for $D$ spatial dimensions, the above set of constraints
generates $N_c=1+D+D(D+1)/2 = (D+1)(D+2)/2$ equations
for the set of $N_c$ Lagrange multipliers
$A,\BB,\CC$ forming the {\it quasi}-invariant
$Q_i=A+\BB\cdot \vv_{i} + \CC:\vv_{i}\vv_{i}$, where the symbol
``$:$'' stands for the tensor scalar product.
These equations are generally non-linear (if the entropy function
is not quadratic in the populations) and consequently
very hard to solve--if solvable at all--analytically 
in order to deliver closed
expressions for the overconstrained local equilibrium as a function of the
hydrodynamic fields. But even this is not the main drawback.
More importantly, such over-constrained equilibria confine
the domain of the phase space where the entropy production
is positive to rather ``thin'' subsets (Karlin and Succi, 1998).
This explains why the Lattice Boltzmann method offers less
stability in such cases than non-isothermal hydrodynamics, where
constructions of the equilibria have been focused mostly on
satisfying the conservation constraints regardless of their origin.

\section{Directions for future research}

To date, there are still several directions
where the Lattice Boltzmann theory calls for
substantial progress and upgrades.
Here, we shall briefly touch upon just two major 
streams of development: 
{\em
\begin{itemize}
\item{} Thermal flows
\item{} Non-ideal fluids
\end{itemize}
}
In a nutshell, the problem is to guarantee a sufficient lattice
symmetry to ensure the correct evolution of {\it both} kinetic
and potential energy, without loosing numerical stability. 
This is fairly non-trivial, since it involves control on
higher-order local kinetic momenta, such as the heat flux, as well
as on non-local information due to potential energy interactions.

As a further prospective area of future development we mention
\begin{itemize}
\item{} {\it Quantum systems}
\end{itemize}
for which much less has been done to date.

\subsection{Thermal LBE's}

Thermal field theories are notoriously hard
(Umezawa, 1993), and Lattice Boltzmann theory is no exception.
To date, isothermal, or better said, {\it athermal} LB
systems\footnote{Since LBE is based on a collection of mono-energetic
beams $\delta(\vv-\vv_i)$ in velocity space, the discrete LB distribution 
function
is more appropriately classified as a zero-temperature system.
We shall nonetheless stick to the more intuitive, if less
rigorous, notation of {\it isothermal} schemes, to indicate that
temperature is not a dynamic variable.} are way better understood
than their thermal counterparts.
The correct treatment of energy degrees of freedom in
a lattice still poses a number of basic questions.

Energy dynamics in Lattice Boltzmann models is accounted for by
enlarging the set of discrete speeds, $26$ being the
number of kinetic moments to be matched by 
the set of discrete speeds.
Early experiments (McNamara, Garcia and Alder, 1995) showed that
thermo-hydrodynamic LBE's exhibit a high degree of
{\it molecular individualism}; errors in higher order
moments seem to penetrate down into the low-order
thermo-hydrodynamic manifold.
Similar type of difficulties have been recognized since long
also in continuum kinetic theory (Lewis, 1967).

Despite significant progress, starting with the work 
of Alexander et al (Alexander, Chen and Sterling, 1993)
thermal LBE's remain less robust than their isothermal counterparts.

Thermal LBE's might be subject to a sort of numerical instability 
similar to the one affecting high-order finite difference
schemes: due to the presence of high-energy
particles, large-set of discrete speeds give rise to 
high-order dispersion relations, which in turn might 
develop spurious branches and unphysical solutions.

This points even more strongly to the importance
of a lattice $H$ theorem.
Again, a qualitative effect of lattice discreteness enters the scene.
In the continuum, the local $H$ theorem
(positive entropy production due to collisions alone)
does {\it not} explicitly involve transition probabilities
from pre- to post-collisional states, and
the detailed balance among different states
are determined only by particle distributions with
no other weighting factors.
As a result, the $H$ theorem is
unaffected by particle advection since entropy is
passively transported along collision-free trajectories.
{\em Consequently the local $H$ theorem automatically implies
the overall $H$ theorem (or ``global $H$ theorem'')}.
In the discrete case, however, {\it the $H$ theorem
does depend explicitly on transition probabilities},
which are functions of local collisional invariants
(Molvig {\em et al.}, 1988; Teixeira, 1992;
Chen, Teixeira and Molvig, 1997; Chen, 1995; Chen, 1997; 
Chen and Teixeira, 2000). This is necessary to ensure that
the resulting local equilibrium has
a suitable form for producing correct thermo-hydrodynamic
equations satisfying Galilean invariance 
(Frisch, Hasslacher and Pomeau, 1986; Frisch {\em et al.}, 1987).
The condition for achieving the correct thermohydrodynamics
requires the equilibrium form given by Chen (1995), namely
an expansion up to $O(u^3)$ of a discrete local Maxwellian:
\begin{equation}
\label{TLEQ}
f_i^{\rm eq} = \rho g_i(T) e^{\epsilon_i/T} exp[- ({\vv}_i - \uu)^2/2T]
\end{equation}
where $T$ is the local hydrodynamic temperature,
\[ \rho (DT + \uu^2)/2 = \sum_i \epsilon_i f_i \]

This form can be realized via explicit lattice gas
particle collisions involving transition probabilities
of $g_i(T)$ (Chen, 1995; Chen, 1997).
On the other hand, as per the discussion above,
relaxation to a given local equilibrium needs not proceed
through explicit collisions.
It can be shown that (\ref{TLEQ}) admits a local $H$ function
of the form:
\begin{equation}
h = \sum_i f_i ln(f_i/g_i)
\end{equation}
The global $H$ function may be simply defined as
\[ H = \sum_{\xx} h({\xx}). \]
Because of its explicit dependence on transition probabilities,
the $H$ value changes during the LBE advection phase, the
result being that as particles hop in discrete steps
from one lattice site to another, the values of the $H$ function
change {\it ahead} of the transition probabilities (which
change during advection).
This hinders translational invariance and sets the stage
for a spontaneous symmetry breaking of the $H$ theorem
with no counterpart in the continuum.
Since the transition probabilities, $g_i$,
are functions of local hydrodynamic temperature, their
value changes from place to place as particles move around the lattice.
The violation of the global $H$ theorem manifests itself whenever
the temperature field acquires a spatial dependence.
On the other hand, though the transition probabilities
are not unity, they become constants in the isothermal
LB models, which explains why the above
problem disappears for isothermal LB models. 
This essential difference between thermal and
isothermal lattice models lies at the heart of the
significantly better stability of the latter.

In summary, aside from low-Mach number expansions in the usual
lattice BGK, isothermal models have been shown to possess an
$H$ theorem (i.e., global $H$ theorem), while thermal models have not.
Proving an $H$ theorem and subsequently constructing
an $H$ theorem obeying the Lattice Boltzmann process for thermal models 
remains an outstanding problem in Lattice Boltzmann research.

\subsection{LBE for non-ideal fluids}

Lattice Boltzmann schemes for multiphase, multicomponent fluid flows are
often heralded as a most promising territory for
LBE research. Indeed, there exist several extensions of the
plain Lattice Boltzmann schemes for hydrodynamics 
which are able to include non-ideal gas effects 
(intermolecular interactions) (Shan and Chen, 1993, 1994
Swift, Osborne and Yeomans, 1995, Luo 1998, 2000).
Most of these models can be cast in the form of a generalized LBE
where the right-hand side is enriched with a 
({\it self-consistent})
source term $F_i$:
\begin{equation} \label{LBGK2}
f_i({\xx}+{\vv}_i,t+1)-f_i({\xx},t)= -\omega[f_i-f_i^{\rm eq}]({\xx},t)
+F_i
\end{equation}
Formally, $F_i$ is the lattice version of the
one-body self-consistent force (vector notation relaxed for simplicity):
\begin{equation}
F(1) f(1) =\int f_{12} (1,2) F_{12} (1,2) d2
\end{equation}
where $f_{12}$ is the two-body distribution function,
$F_{12}$ the two-body force, and $1,2$ stand for six-dimensional
phase-space coordinates.
Since the two-body distribution is likely to be computationally intractable,
Lattice Boltzmann research moved in the direction of a
{\it dynamic lattice density functional theory}
(Hansen and Mc Donald, 1986), in which the effective
force $F(1)$ is represented by a semi-empirical functional
of the macroscopic state of the system, namely density, flow fields 
and their derivatives:
\begin{equation}
F(1) = - \nabla \Psi [\rho, {\uu}, ...]
\end{equation}

These schemes have been validated for a series of test cases
and complex flows applications, yielding fairly
interesting results (as an illustration, see Figure 7).

However, none of them has been proved to admit an
underlying $H$ theorem (Rothman, Keller and Gunstensen, 1991; 
Shan and Chen, 1993; Shan and Chen, 1994;
Swift, Orlandini and Yeomans, 1995).
This is in part due to the fact that equilibrium distribution
functions depend on non-local properties. In other words,
non-local information determines the transition probabilities.
Consequently, similar to the problem encountered in the thermal
LB models, advection changes globally defined quantities
such as a $H$ function. 
Another major theoretical issue is the nature of the hydrodynamic
limit near thin-interfaces where the low-Knudsen assumption
behind LBE is seriously challenged (a similar issue arises in the
formulation of LBE-based renormalization group treatments of fluid
turbulence (Chen, Succi, Orszag, 1999)). 

Finally, as indicated earlier in this paper, 
the properties of the $H$ theorem in the presence of a self-consistent force,
are much less understood even in the continuum, 
Even though a global entropy production in the sense
of decreasing $H$ still exists in continuum Boltzmann,
the issue of convexity and uniqueness of a minimum point becomes moot
because the local condition $dH/dt<0$ no longer guarantees
that collisional dynamics will attain a {\it global} minimum.
Besides serving the purpose of formulating better Lattice 
Boltzmann models, we believe this can be regarded as
a fundamental research topic of general interest in
non-equilibrium statistical mechanics (Lebowitz, 1999).

\subsection{LBE for quantum systems}

To date, Lattice Boltzmann research is largely dominated
by non-quantum physics applications.
Nonetheless, LBE can and has indeed been extended to (simple) cases
of quantum mechanical motion, typically in the form
of the non-relativistic Schroedinger equation (Succi and Benzi, 1993, 
Succi 1996, Meyer 1997, Boghosian and Taylor 1997). 
The stepping stone to the quantum Lattice Boltzmann equation is 
the identification of the four-spinor wave function $\Psi_i$ 
obeying the quantum relativistic Dirac equation with
a {\it complex} discrete distribution function.
By writing the Dirac equation as a complex
Lattice Boltzmann equation, it can be shown that the 
non-relativistic limit taking the Dirac equation
into the Schroedinger equation is formally analogous 
to the adiabatic limit yielding hydrodynamics
from the Lattice Boltzmann equation.

The proof proceeds by writing the Dirac equation (in one
dimension for simplicity) for a relativistic particle
of rest mass $m$ as two coupled LBE's for a pair
of complex upward and downward propagating spinors
$u(z,t)$ and $d(z,t)$ respectively:
\begin{eqnarray}
\partial_t u + \partial_z u = \omega_c d\\\nonumber
\partial_t d - \partial_z d = -\omega_c u
\end{eqnarray}
where $\omega_c = m c^2/\hbar$ is the Compton frequency of 
a material particle.
By applying the unitary transformation:
\[
\phi ^{\pm} = \frac{1}{\sqrt 2} (u \pm id) \; e^{i \omega_c t}
\]
the above set of equations takes
the following hydrodynamic form:
\begin{eqnarray}
\partial_t \phi^+ + \partial_z \phi^- = 0\\\nonumber
\partial_t \phi^- - \partial_z \phi^+ = 2 i \omega_c u
\end{eqnarray}
From these equations it is apparent that the ``slow'' 
(hydrodynamic) symmetric mode is conserved, whereas the 
``fast'' antisymmetric mode oscillates
with a doubled frequency $2 \omega_c$.
It is now readily checked that the adiabatic assumption:
\[
|\partial_t \phi^-| << | 2 i \omega_c \phi^-|
\]
delivers precisely the Schroedinger equation 
for a free-particle of mass $m$. 
Since the adiabatic approximation involves an imaginary
frequency, unlike the classical case,
the hydrodynamic branch (i.e., the Schroedinger equation) 
remains time-reversible.
This reversibility is somehow weaker than for the 
Dirac equation since it only applies to real time, whereas
the Dirac equation is reversible in both real and imaginary time. 
Interesting connections to a quantum $H$-theorem might arise
in the framework of the $N$-body quantum Lattice
Boltzmann equation, a totally unexplored issue to the best of our knowledge.

The quantum LBE can be turned into a practical numerical scheme 
for the non-relativistic Schroedinger equation as well as
for relativistic wave mechanics (Succi, 2002).
It might also be suitable to quantum computing paradigms.

Saying that quantum LBE is a major issue of future Lattice
Boltzmann research is probably an overstatement. Still,
owing to the growing interest in quantum computing, 
the subject might be worth some attention in the years to come.

\section{Conclusions}

These considerations complete the description of the Lattice
Boltzmann method for isothermal Navier-Stokes equations
as a self-contained kinetic theory admitting a proper $H$ theorem.
The corresponding theory for fully thermo-hydrodynamic LBE's
is still in its infancy. 
Not only is further understanding of this subject
crucial to the formulation of more stable numerical
algorithms, but it might also stimulate 
new insights into fundamental physical questions involving 
temperature dynamics and non-local intermolecular 
interactions in discrete dynamical systems.

\section{Acknowledgments}

The authors are thankful to V. Boffi, B. Boghosian, A. N. Gorban, 
L. Luo,  H. C.  \"Ottinger for many valuable discussions.

%Figure 1
\begin{figure}
    \includegraphics[width=0.50\textwidth]{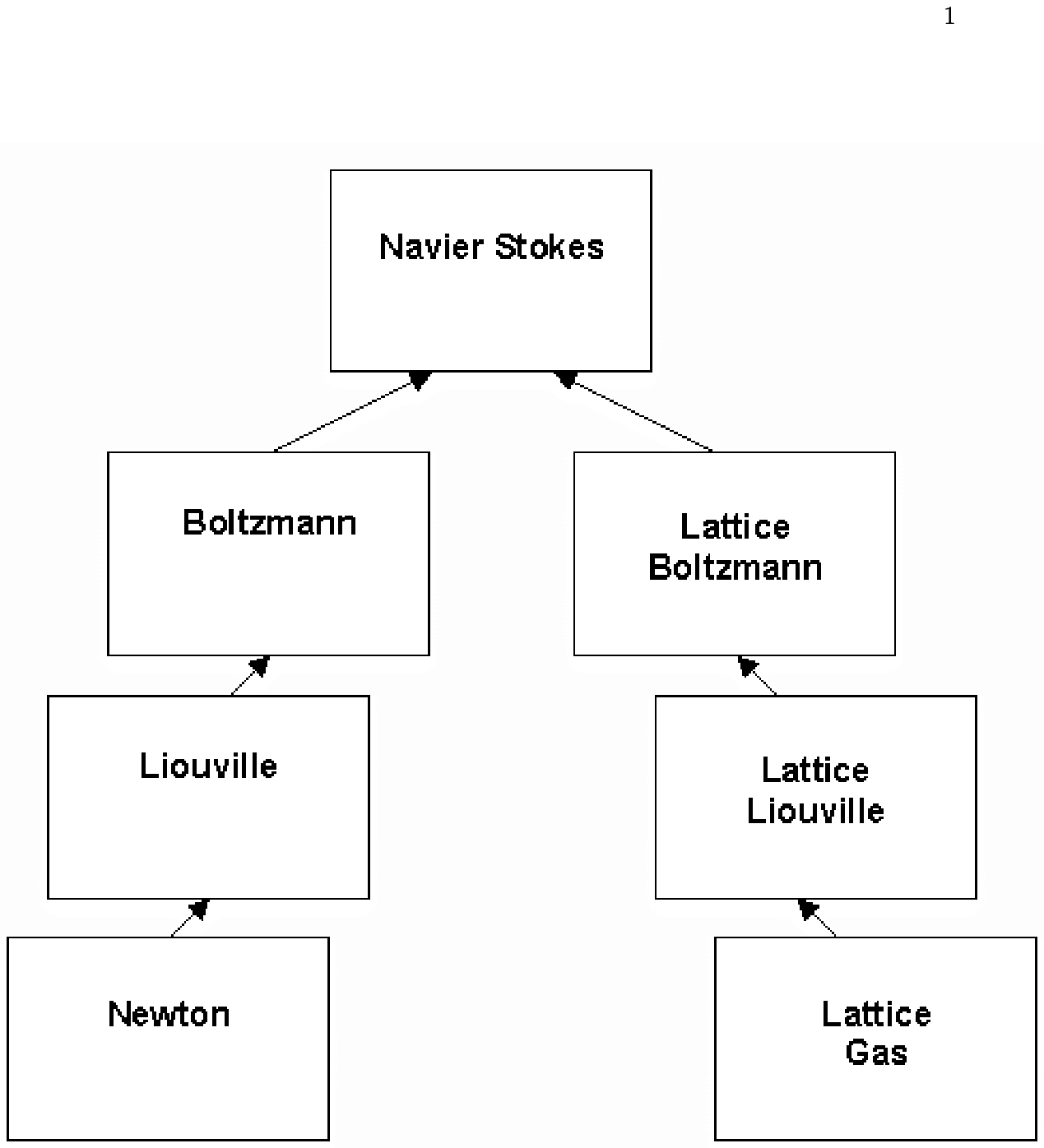}
\caption{\label{Fig1}The BBGKY hierarchy and its lattice analogue.
Lack of microscopic detail becomes less and less relevant
as one proceeds upwards along the hierarchy
(from Succi, (2001)).}
\end{figure}

%Figure 2
\begin{figure}
    \includegraphics[width=0.50\textwidth]{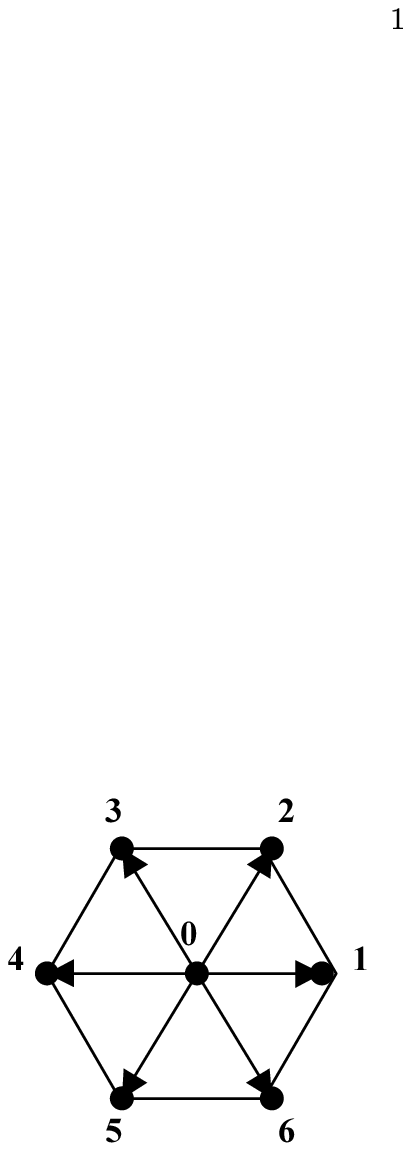}
\caption{\label{Fig2}The hexagonal lattice of the Frisch-Hasslacher-Pomeau 
cellular automaton. Particles move along the six discrete links
and meet at lattice nodes where they interact according to mass
and momentum-conserving collision rules.}
\end{figure}

%Figure 3
\begin{figure}
    \includegraphics[width=0.50\textwidth]{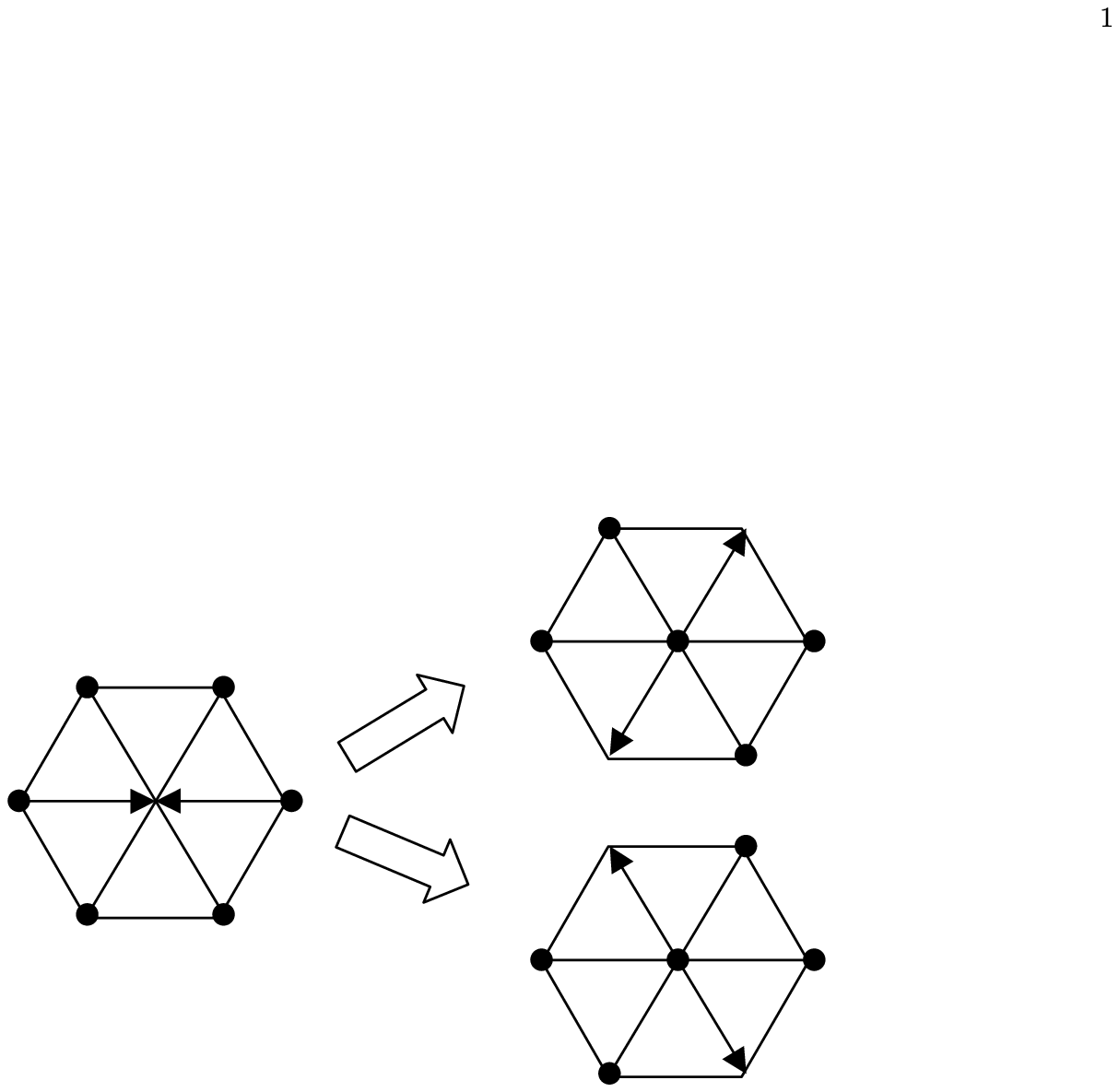}
\caption{\label{Fig3}A typical collision in the Frisch-Hasslacher-Pomeau cellular
automaton. Both post-collisional outcomes (right) are equally probable. 
The left hand hexagon shows two particles initially at 9 o'clock and 3 o'clock
moving in one time step to the central site. In the next time
step, indicated by the hexagons on the right, their collision
(conserving energy and momentum) results in the particles occupying
the sites at 11 o'clock and 5 o'clock or 1 o'clock and 7
o'clock. They could return to their initial locations at 9
o'clock and 3 o'clock but this event is ruled out since it
would not produce any observable effect since the particles
are indistinguishable.
}
\end{figure}

%Figure 4: 
\begin{figure}
 \includegraphics[width=0.50\textwidth]{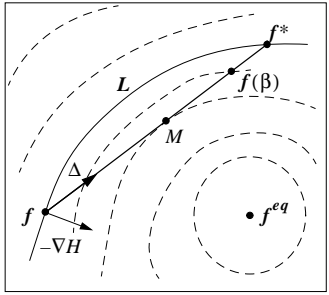}
\caption{\label{Fig4}Collisional relaxation procedure. Curves represent
entropy levels, surrounding the local equilibrium $f^{\rm eq}$.
The solid curve $L$ is the entropy level with
the value $H(f)=H(f^*)$, where
$f$ is the initial, and $f^*$ is
the conjugate population. The vector $\bDelta$ represents
the collision integral, the sharp angle between $\bDelta$ and
the vector $-\bnabla H$ reflects the entropy production inequality.
The point $M$ is the minimum entropy state
on the segment $[f,f^*]$ 
The result of the collision update is represented by the point
$f(\beta)$.
The choice of $\beta$ shown corresponds to the ``overrelaxation'':
$H(f(\beta))>H(M)$ but $H(f(\beta))<H(f)$. The particular
case of the BGK collision (not shown)
would be represented by a vector $\bDelta_{\rm BGK}$, pointing 
from $f$ towards $f^{\rm eq}$, in which case $M=f^{\rm eq}$.}
\end{figure}

%Figure 5
\begin{figure}
 \includegraphics[width=0.50\textwidth]{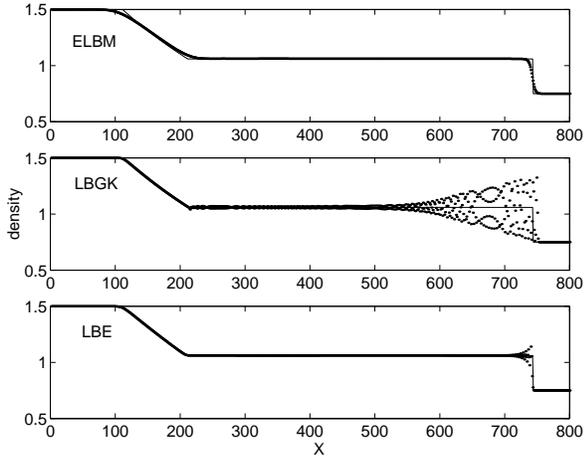}
\caption{\label{Fig5}Evolution of a one-dimensional front in a shock tube.
Density profile (dimensionless lattice units) is shown at 
$t=500$ for viscosity $\nu=3.3333\times 10^{-2}$. 
The figure compares three Lattice Boltzmann algorithms on
the lattice with $800$ nodes.
Thin line: Exact solution at zero viscosity. 
Solid circles: Simulation.
ELBM: The Lattice Boltzmann method based on the entropy function (from
Ansumali and Karlin (2000)).
LBGK: The Lattice BGK algorithm based on the polynomial ansatz
of Qian, d'Humi\`{e}res and Lallemand (1992).
LBE: The Lattice Boltzmann model 
of Qian, d'Humi\`{e}res and Lallemand (1991).
The value of viscosity is taken close to the 
instability of the LBE.}
\end{figure}

%Figure 6
\begin{figure}
 \includegraphics[width=0.50\textwidth]{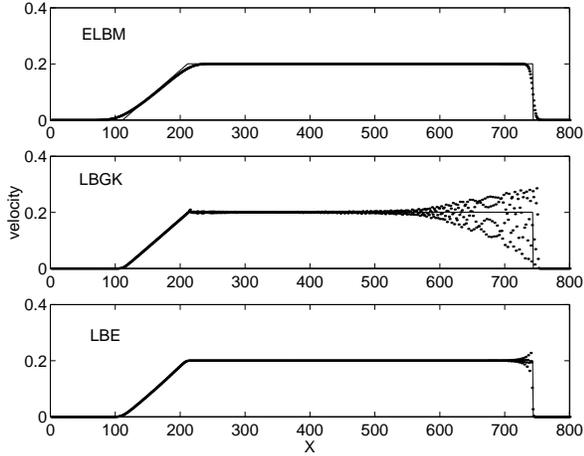}
\caption{\label{Fig6}Velocity profile in the one-dimensional shock tube
benchmark. Simulation setup and notation are the same
as in Figure 5}
\end{figure}

%Figure 7
\begin{figure}
 \includegraphics[width=0.50\textwidth]{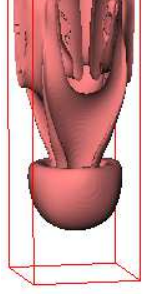}
\caption{\label{Fig7}A typical fluid interface exhibiting a 
single-mode Rayleigh-Taylor instability associated
with a heavy fluid sitting on top of a light one.
The density ratio of heavy and light
fluids is $3$. $Re = (gL)^{1/2} L / \nu = 1024$, 
$t / T_0 = 3.5$ ($T_0 = L/(gL)^{1/2}$), $g$ being the
gravitational acceleration driving the instability and
$L$ the box size. 
Periodic boundary conditions are used in horizontal
directions and no slip boundary conditions are used in vertical
directions (from Zhang, Chen and Doolen (1999), courtesy of Raoyang Zhang).}
\end{figure}

\end{document}